\documentclass[aps,prb,final,twocolumn]{revtex4}   
\usepackage{graphicx}

\usepackage{amsmath}    
\usepackage{amssymb}    
\usepackage{graphicx}   
\usepackage{array}
\usepackage{epstopdf}

\begin{document}
\title{Interatomic potentials for mixed oxide (MOX) nuclear fuels }
\author{Pratyush Tiwary}
 \affiliation{Division of Engineering and Applied Sciences, California Institute of Technology, Pasadena, California 91125, USA}
\author{Axel van de Walle}
\affiliation{Division of Engineering and Applied Sciences, California Institute of Technology, Pasadena, California 91125, USA}
\author{Byoungseon Jeon}
\affiliation{Department of Applied Science, University of California, Davis, California 95616, USA}
\author{Niels Gr{\o}nbech-Jensen}
\affiliation{Department of Applied Science, University of California, Davis, California 95616, USA}
\date{\today}

\begin{abstract}
We extend our recently developed interatomic potentials for UO$_2$ to the mixed oxide fuel system (U,Pu,Np)O$_2$. We do so by fitting against an extensive database of \textit{ab initio} results as well as to experimental measurements. The applicability of these interactions to a variety of mixed environments beyond the fitting domain is also assessed. The employed formalism makes these potentials applicable across all interatomic distances without the need for any ambiguous splining to the well-established short-range Ziegler-Biersack-Littmark universal pair potential.  We therefore expect these to be reliable potentials for carrying out
damage simulations (and Molecular Dynamics simulations in general) in nuclear fuels of varying compositions for all relevant
atomic collision energies.
\end{abstract}
\maketitle
The interest in using Mixed Oxide (MOX) fuels comprising (U,Pu,MA)O$_2$ (where MA =  Np, Am and Cm) in fast breeder and transmutation reactors is ever increasing. Since this complex fuel experiences a high burn-up ratio with large quantities of fission products and materials defects, it becomes crucial to understand the evolution and statistics of atomic displacement cascades due to high energy radiation that the material faces\cite{devan1}. Classical Molecular Dynamics (MD) with its ability to simulate fairly long length scales, though still retaining the fine atomic structure of the material, is ideally suited for such studies. However, the complexity of the interatomic interactions for radiation damage simulations cannot be fully represented by simple classical forms due to the disparate scales of energies involved. Interactions corresponding to equilibrium conditions are traditionally found by fitting to a variety of thermodynamic data; while for description of the short-range behavior, the Ziegler-Biersack-Littmark (ZBL) universal pair potential\cite{zbl} developed in the 1980s is well-accepted. These two ``pieces'' then need to be smoothly connected via somewhat arbitrarily applied splines. We recently proposed a methodology for developing interatomic potentials that is valid for all interatomic separations, without the need for any ambiguous splines\cite{Tiwary}.  In this article, we apply this formalism to a more general case of MOX nuclear fuels of varying composition. In addition to capturing high temperature thermodynamic properties, as done by available potentials\cite{prevpot1,prevpot2,prevpot3}, we also incorporate correct treatment of point defects. Created due to irradiation, these are critical for the understanding of a variety of phenomena such as fuel swelling, fission gas release and burn-up structure formation \cite{morelon,brutzel}.
A key test of any developed energy surface lies in its ability to adequately represent systems/configurations that were not included in the fitting procedure.\cite{avdw:maps} We here fit the potential parameters to \textit{ab initio} and experimental data for the oxides PuO$_{2}$ and NpO$_{2}$, and then check for their transferability by comparing against \textit{ab initio} data for (U$_x$Pu$_{1-x}$)O$_2$ and (U$_x$Np$_{1-x}$)O$_2$ configurations that were not included in the fit.

In the present study we employ the generalized potential formalism \cite{Tiwary} that behaves correctly in both short-range and long-range limits. The only component in this potential that remains to be determined is a correction term for intermediate distances associated with chemical bonding. We find this correction term by fitting to an extensive database of generalized gradient approximation GGA+\textit{U ab initio} calculations\cite{anisimov} on PuO$_{2}$ and NpO$_{2}$. The potential's applicability in a mixed environment pertinent to MOX fuels is further verified by testing against GGA+\textit{U} data for (U$_x$Pu$_{1-x}$)O$_2$ and (U$_x$Np$_{1-x}$)O$_2$. GGA+\textit{U} is known to provide electronic and magnetic behaviors of the actinide oxides\cite{moore} that are consistent with experiments. In this approximation, the spin-polarized GGA potential is supplemented by a Hubbard-type term to account for the localized and strongly correlated 5\textit{f} electrons. Our database comprises results obtained from GGA+\textit{U} calculations with the projector augmented-wave method and collinear antiferromagnetic moments as implemented in the VASP package\cite{vasp}. Dudarev's rotationally invariant approach\cite{dudarev1,dudarev2} to GGA+\textit{U} is employed wherein the parameter U-J is set to 3.99, 3.25 and 3.40 for U, Pu and Np respectively\cite{geng1,geng2,andersson}. These are the generally accepted values for reproducing the correct band structures of the corresponding oxides. Energy cutoff for the plane waves was kept at 400 eV. Since GGA+\textit{U} overestimates the lattice parameter, a common scaling factor (same as that used\cite{Tiwary} for UO$_2$) was employed to get experimentally correct lattice parameters. The \textit{ab initio} database so obtained for fitting comprises:
\begin{enumerate}
\item Isochoric relaxed runs on a 12 atom unit cell, which was isometrically contracted and expanded by various amounts (i.e., equation of state calculations wherein each data point was calculated under the constraint of constant cell volume) and for which an 8$\times$8$\times$8 \textit{k}-point grid was taken after ascertaining \textit{k}-point convergence. Ionic relaxations were carried out until residual forces were less than 0.01 eV/{\AA}.
\item Static (i.e., no ionic relaxation) runs on a 96 atom 2$\times$2$\times$2 supercell in which one atom at a time (O or Pu or Np) was perturbed from its equilibrium position by varying distances (on the order of 1 {\AA} or less from the equilibrium positions) in different directions. Sampling of the gamma point only was found to be satisfactorily accurate for this.
\item A 96 atom 2$\times$2$\times$2 supercell for the formation energies of stoichiometric defects, namely, Oxygen Frenkel pair, Neptunium Frenkel pair and Plutonium Frenkel pair. Several vacancy-interstitial distances were considered to ascertain the separation between these corresponding to the minimum defect formation energy (excluding the case of nearest neighbor distances, which was found to lead to vacancy-interstitial recombination). Correct prediction of these energies has been given great importance in generating interatomic potentials for cascade simulations in UO$_{2}$ \cite{devan1,devan2,morelon,brutzel,govers1,govers2}.
\end{enumerate}
A total of approximately 50 \textit{ab initio} configurations were thus used in the fitting. Note that in the above calculations, any interactions between atoms and their periodic images do not systematically bias the fit of the potentials because the same supercell geometry is used in both the \textit{ab initio} and the empirical potential energy calculations.

The \textit{ab initio} database employed for validation and for testing transferability includes equation of state runs similar to those in the fitting database, for oxides of U$_{31}$Pu, U$_{30}$Pu${_2}$, U$_{31}$Np and U$_{30}$Np${_2}$, each with 64 Oxygens. These data points were not included in the fit itself and were used only after the fitting was complete for checking the robustness of the potentials with respect to use in mixed environments.

\begin{figure}[htp]
\begin{center}
 \includegraphics[width=90mm]{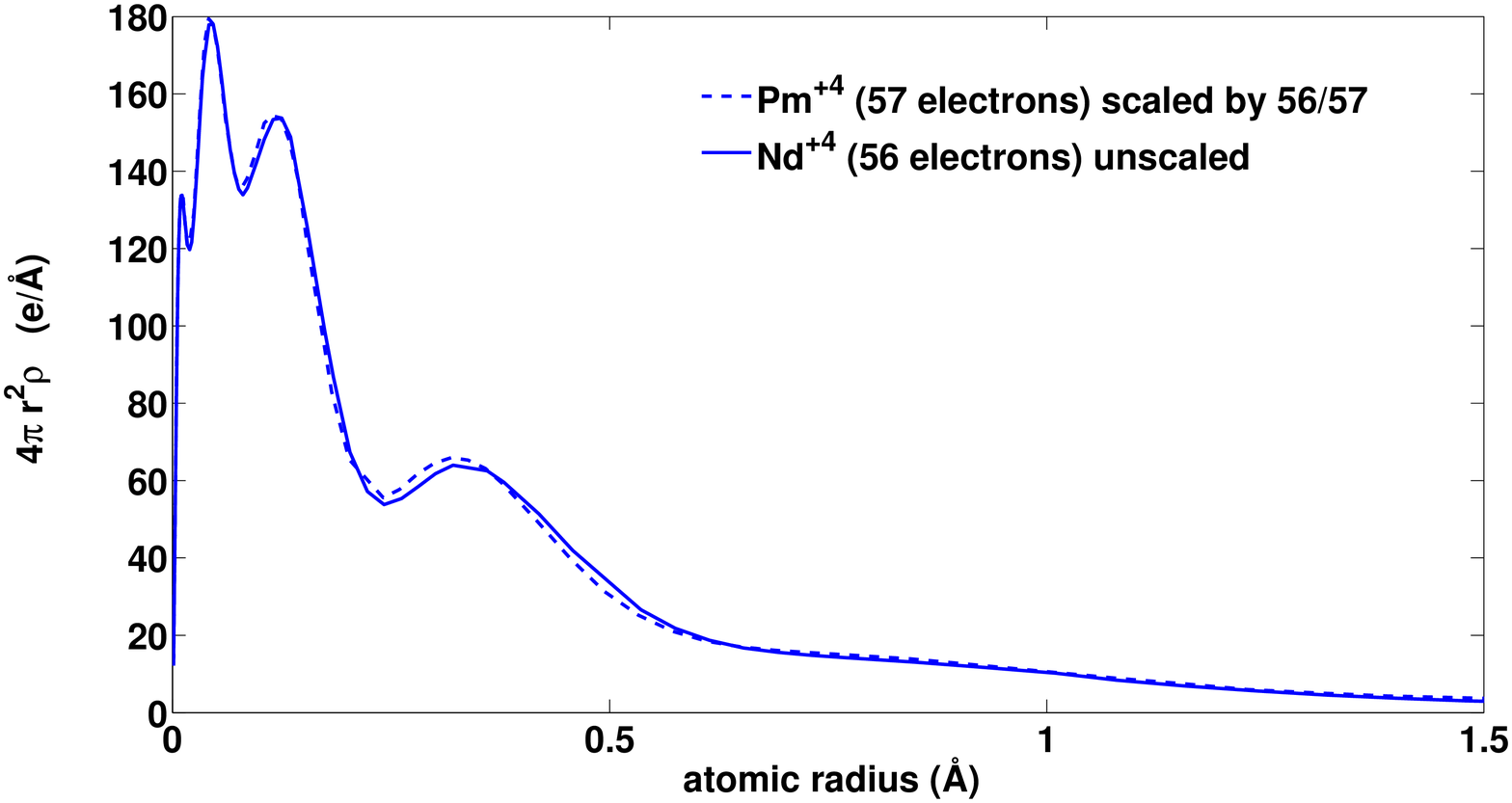}
 \end{center} 
 \caption[2]
{Test of approximation-validity of f$_{PuPu}$ = (90/88)f$_{UU}$ and f$_{NpNp}$ = (89/88)f$_{UU}$ by looking at the applicability of similar relations for cations of members of the previous row of the periodic table with similar shell structure viz. Pm and Nd. Dashed line denotes the result from this approximation while solid line is the actual charge density \cite{zbl}  for Nd$^{+4}$ . }
\label{fig:f1}
\end{figure}

\begin{figure}[htp]
\centering
 \includegraphics[width=90mm]{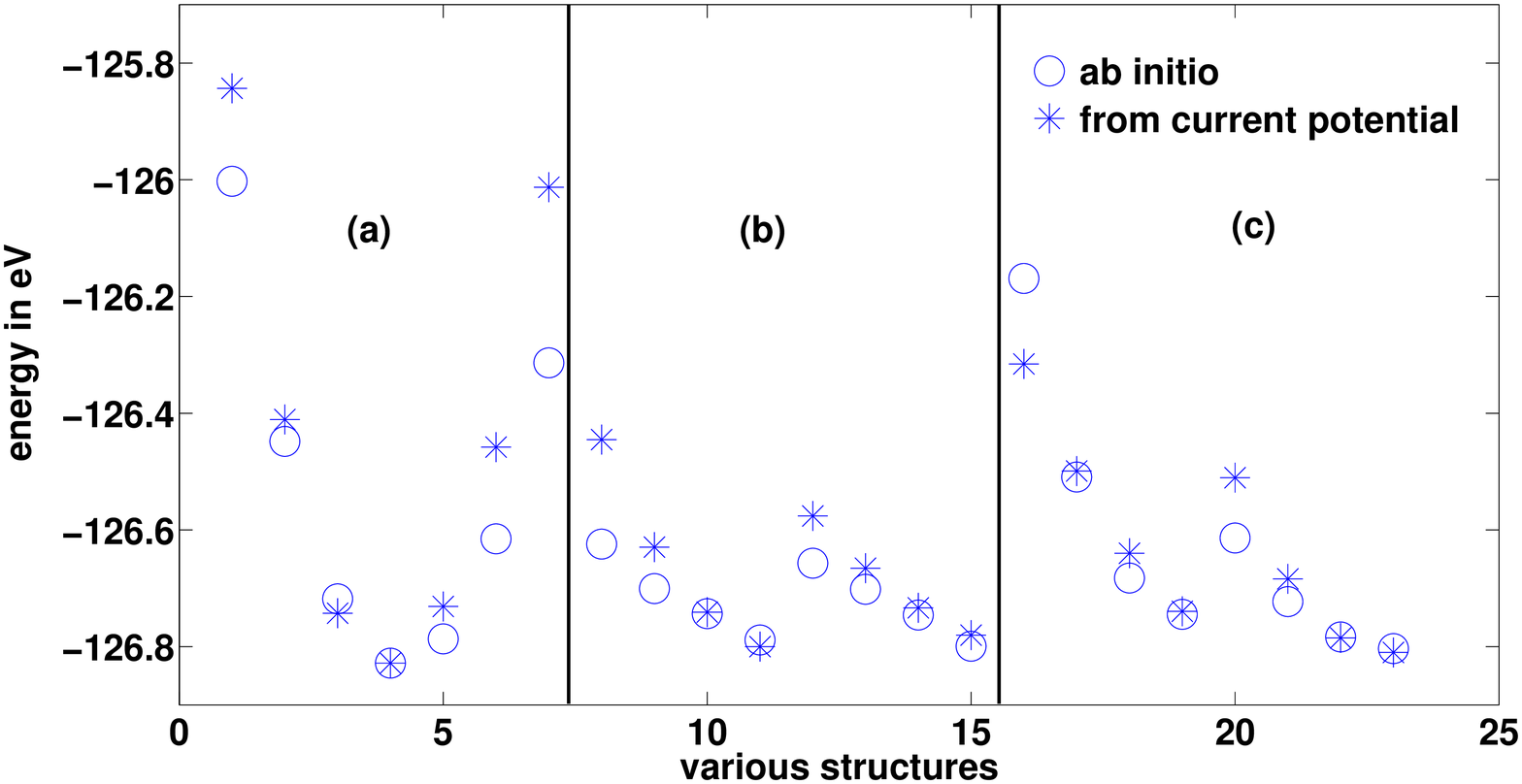}
  \caption[2]
{Quality of fit from our fitted potential (asterisks) for various \textit{ab initio} energies (circles) for PuO$_2$ :(a) equation of state (b) oxygen atom perturbation (c) plutonium atom perturbation. For each of oxygen and plutonium, the first four perturbations are along $\langle100\rangle$ direction while the second four are along $\langle110\rangle$ direction. The perturbations are on the order of 1 {\AA} or lower from the equilibrium positions.}
\label{fig:f2}
\end{figure}

\begin{figure}[htp]
\centering
 \includegraphics[width=90mm]{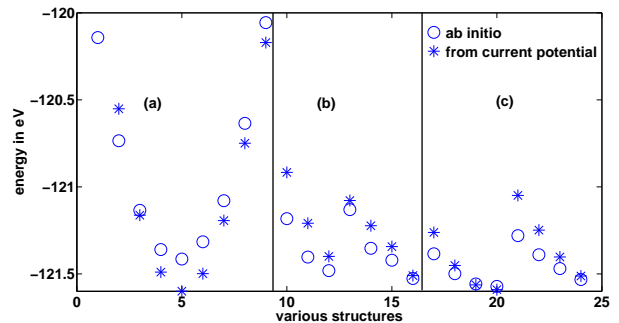}
  \caption[3]
{Quality of fit from our fitted potential (asterisks) for various \textit{ab initio} energies (circles) for NpO$_2$ :(a) equation of state (b) oxygen atom perturbation (c) neptunium atom perturbation. For each of oxygen and neptunium, the first four perturbations are along $\langle100\rangle$ direction while the second four are along $\langle110\rangle$ direction. The perturbations are on the order of 1 {\AA} or lower from the equilibrium positions. }
\label{fig:f3}
\end{figure}

In addition to the \textit{ab initio} data, we also included experimental thermal expansion behavior\cite{latparam} of PuO$_{2}$ and NpO$_{2}$ in the fit. We found that including experimental thermal expansion data (which is readily available) is a very effective way to ensure reasonable thermal expansion behavior in this system. To make the calculation of high temperature lattice parameters computationally tractable during the fitting procedure, we employed the \textit{quasiharmonic approximation} (QHA)\cite{qha}, in which atoms are treated as pure harmonic oscillators whose frequencies depend on the cell volume. The so-called \textit{zero static internal stress approximation} (ZSISA) \cite{zsisa} to QHA, as implemented in the package GULP, was used\cite{gulp}. QHA involves a full relaxation with respect to external (cell parameters) and internal (atom positions within the cell) coordinates. ZSISA ignores the dependence on internal coordinates of the vibrational part of the free energy.  We found that for the materials studied and potential forms used in this communication, the lattice parameter through NPT (constant Number, Pressure, Temperature) MD was slightly lower than that through ZSISA. As such, an empirical adjustment to the ZSISA lattice parameter had to be included in the fitting. Thus, several independent fits were done using ZSISA lattice parameter values equal to the experimental lattice parameter multiplied by $\eta$, with $\eta$ varying between 1 and 1.01. NPT MD was carried out with these potentials (details of MD provided later) to find the $\eta$ that led to MD values matching the experimental data the best. We found that $\eta$ equals 1.0006 and 1.0008 for PuO$_2$ and NpO$_2$ respectively, for a best match in the least squares sense between experimental and NPT MD lattice parameters.

\begin{figure}[htp]
\centering
 \includegraphics[width=90mm]{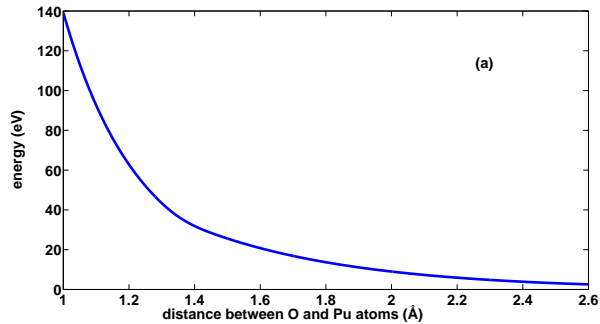}
 \includegraphics[width=90mm]{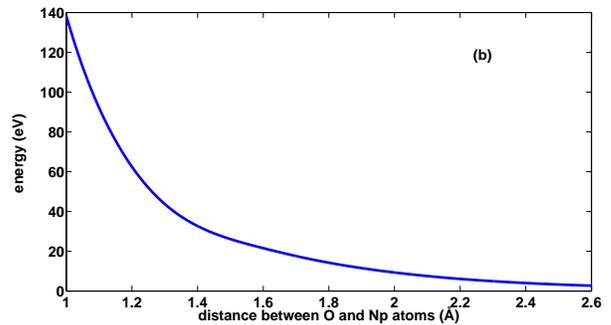}
  \caption[4]
{(a) fitted O-Pu interaction (b) fitted O-Np interaction}
\label{fig:f4}
\end{figure}

\begin{figure}[htp]
\centering
 \includegraphics[width=90mm]{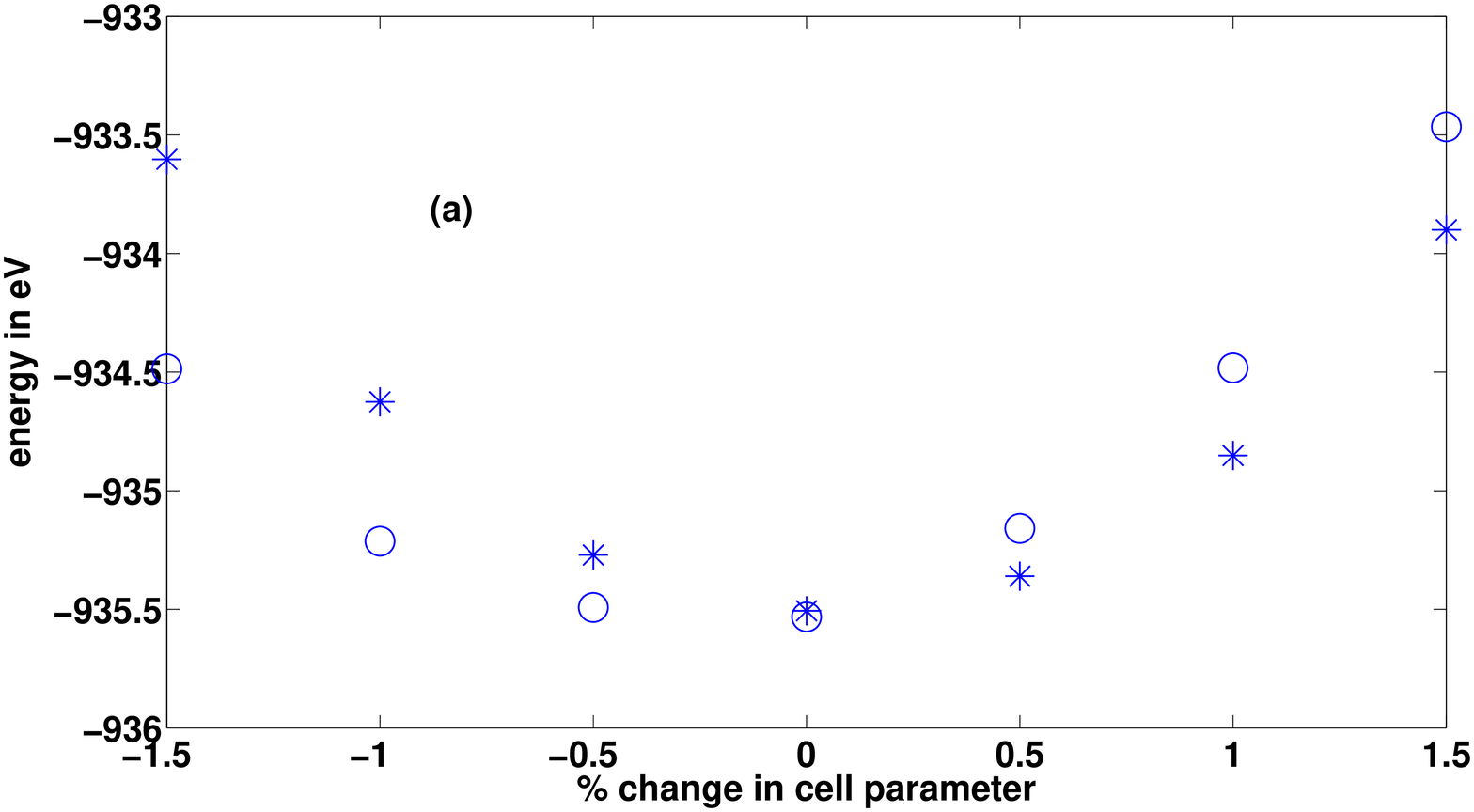}
 \includegraphics[width=90mm]{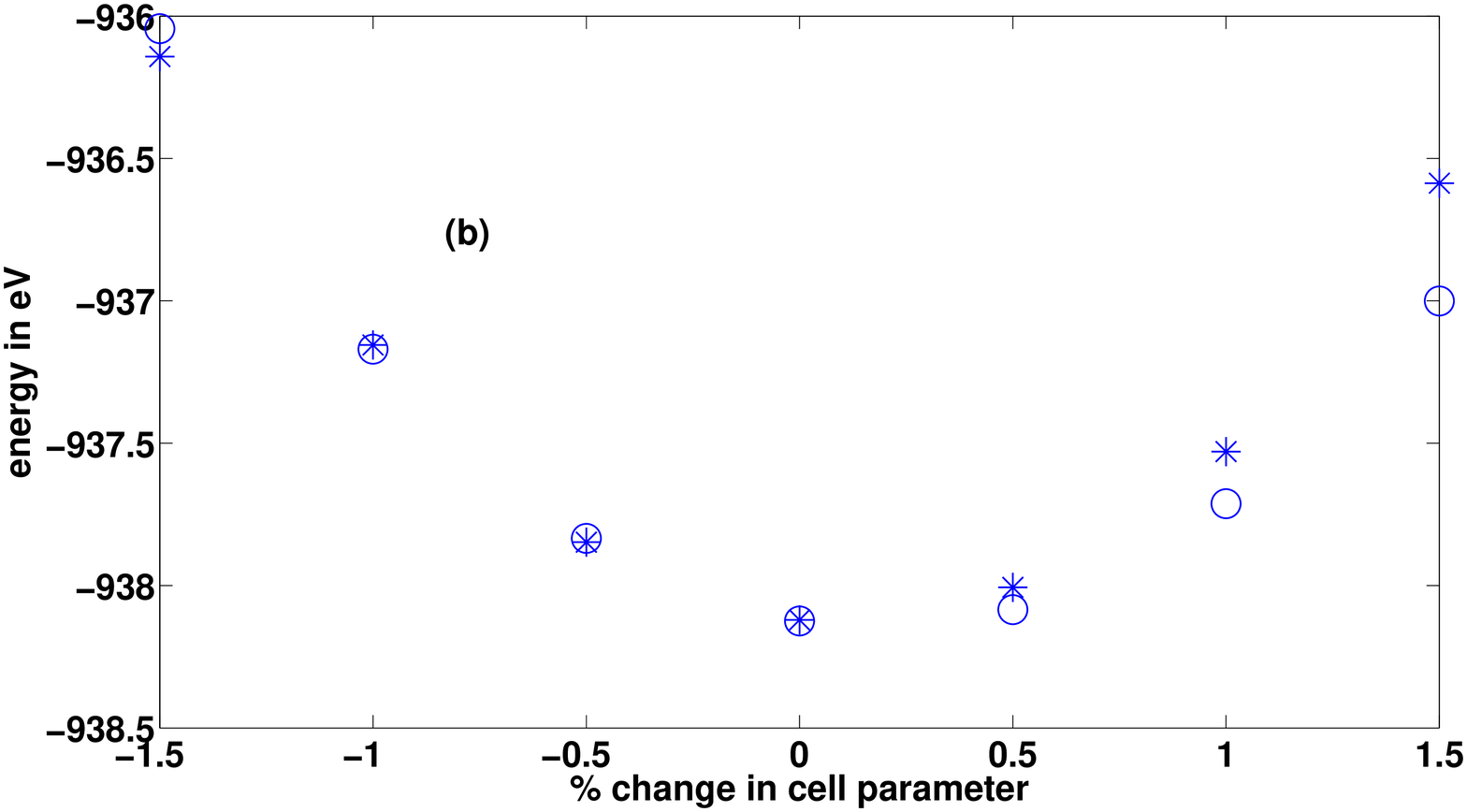}
  \caption[5]
{Equation of state for (a) U$_{31}$PuO$_{64}$ and (b) U$_{30}$Pu${_2}$O$_{64}$. Circles denote \textit{ab initio} data while asterisks are the values predicted (not fitted) with current potential.}
\label{fig:f5}
\end{figure}

\begin{figure}[htp]
\centering
 \includegraphics[width=90mm]{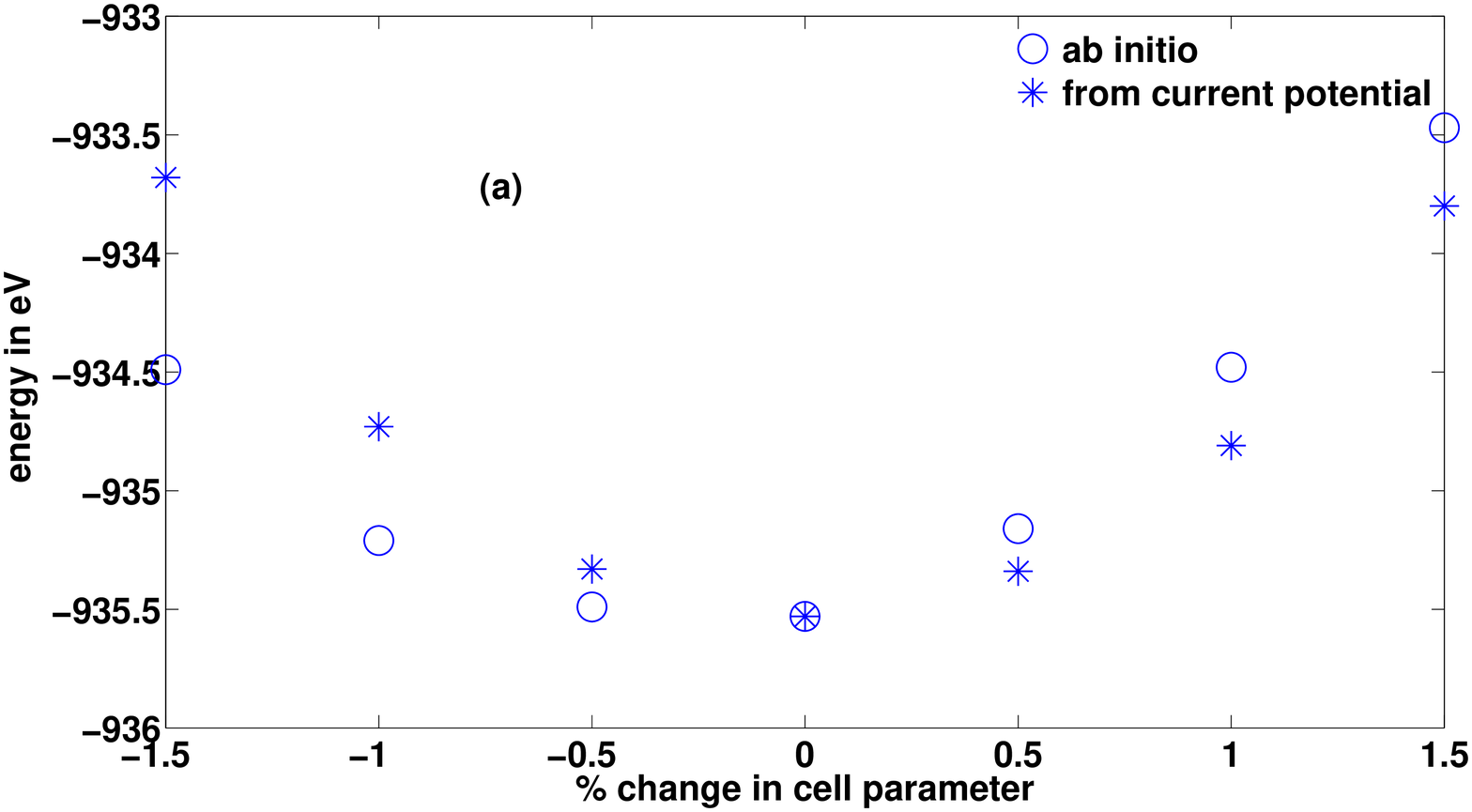}
 \includegraphics[width=90mm]{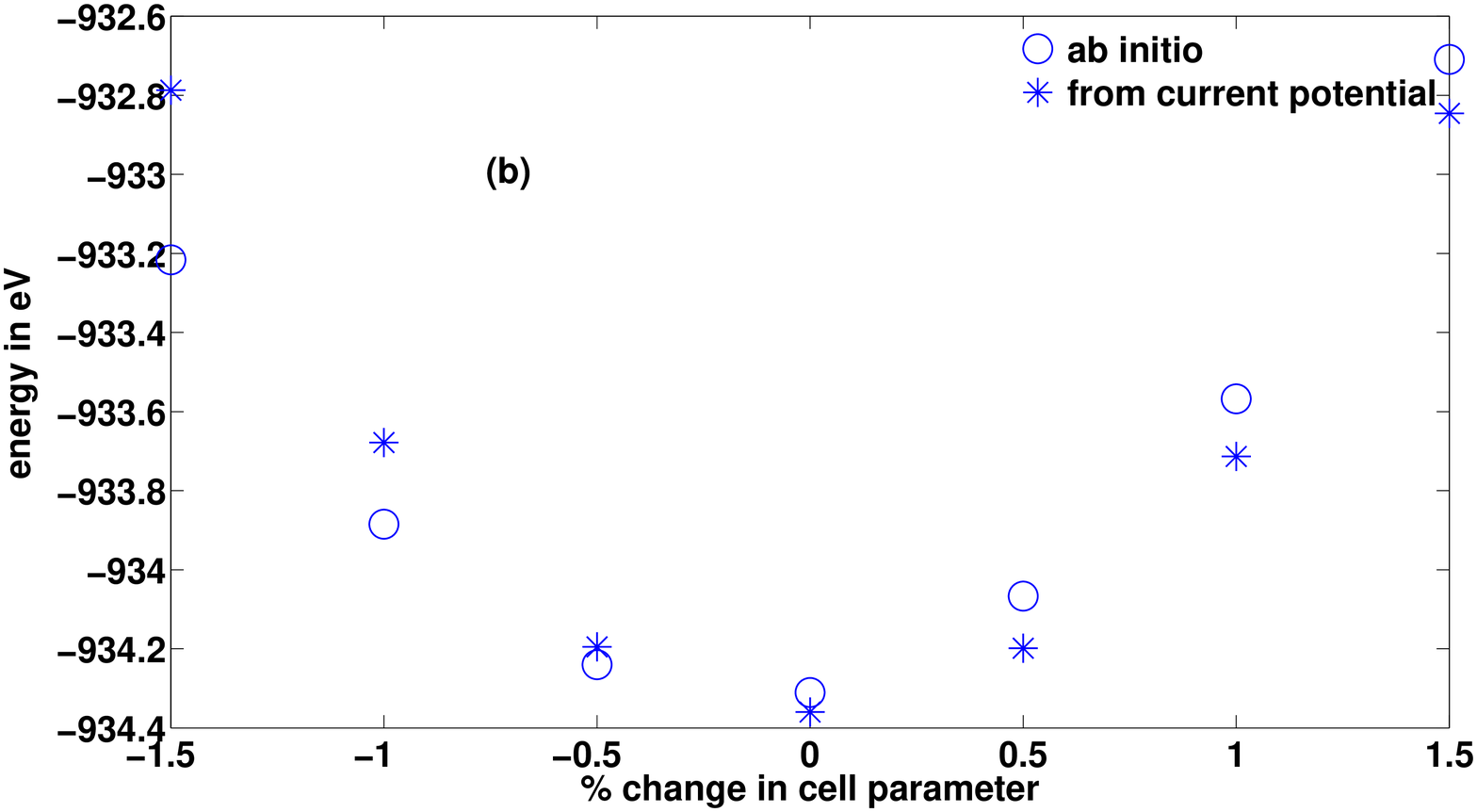}
  \caption[6]
{Equation of state for (a) U$_{31}$NpO$_{64}$ and (b) U$_{30}$Np${_2}$O$_{64}$. Circles denote \textit{ab initio} data while asterisks are the values predicted (not fitted) with current potential.}
\label{fig:f6}
\end{figure}

The potential forms thus used for fitting to the \textit{ab initio} and experimental data are similar to that proposed previously \cite{Tiwary}, and are summarized below for Pu-Pu and Pu-O interactions (with similar forms for other interactions):
\begin{widetext}
\begin{equation}
\label{eq:eq1}
	 V_{PuPu}(r) = ZBL_{90,90} (r) + {{(4)(4)e^2}\over{4\pi\epsilon_0r}} + {{8e^2}\over{4\pi\epsilon_0}}[{90\over r}-{4\pi \over{e}}f_{PuPu}(r)]   \;\;\; \forall \;\;\; 0 < r 
	\end{equation}
\begin{eqnarray}
	V_{OPu}(r) &=& {{(-2)(4)e^2}\over{4\pi\epsilon_0r}}
	 +  \left\{ 
\begin{array}{l l}
  ZBL_{90,10} (r) + {{4e^2}\over{4\pi\epsilon_0}}[{\frac{10}{r}}-{4\pi \over{e}} f_{OO}(r)]  - {{2e^2}\over{4\pi\epsilon_0}}[{\frac{90}{r}}-{4\pi \over{e}} f_{PuPu}(r)] & \quad {0 < r \leq r_1 }\\
    5^{th}\; order \;polynomial & \quad {r_1 < r \leq r_2}\\
     A \exp(-r/\rho)-{B/r^6} + (r-r_3)^{2}(Cr^3 + Dr^2) & \quad {r_2 < r \leq r_3}\\
  A \exp(-r/\rho)-{B/r^6} & \quad {r_3 < r }\\ \end{array} \right. \nonumber 
  \end{eqnarray}
\end{widetext}

\begin{widetext}
\begin{table}[htp]
\caption{ Defect energy comparisons}
\begin{tabular}{ c | m{3cm} | m{3cm} | m{3cm} }
\hline\hline

\hline
       &
       \textit{ab initio} \newline (Current work) &
       Potential \newline (Current work)  &
       Potential \newline (Previous works\cite{prevpot1,prevpot3}) \\
       
       \hline

            O Frenkel pair formation energy in PuO$_2$ (eV) &
       3.9 &
      4.9 &
      7.0 \\
      \hline
       O Frenkel pair formation energy in NpO$_2$ (eV)  &
       4.5 &
      5.8 &
      10.0 \\
      \hline
        Pu Frenkel pair formation energy (eV) &
       11.9 &
      24 &
      17 \\
       \hline
       Np Frenkel pair formation energy (eV) &
       12.2 &
       26.7 &
      17.5 \\
\end{tabular}
\label {table:defect}
\end{table}
\end{widetext}

The UO$_2$ family of interactions is kept the same as in Ref. \onlinecite{Tiwary}. Here $ZBL_{Z_1+q_1,Z_2+q_2} (r)$ denotes the ZBL form of interaction between two neutral atoms having atomic numbers ${Z_1+q_1}$ and ${Z_2+q_2}$, but using the screening length for $Z_1$ and $Z_2$, as explained in Ref. \onlinecite{Tiwary}. The functions \textit{f} in the above are related to the charge densities of the respective atoms. Detailed coefficients of f$_{OO}$ and f$_{UU}$ can be found in Ref. \onlinecite{Tiwary}, while f$_{PuPu}$ and f$_{NpNp}$ can be calculated from the relations f$_{PuPu}$ = (90/88)f$_{UU}$ and f$_{NpNp}$ = (89/88)f$_{UU}$. This was needed since Np$^{+4}$ and Pu$^{+4}$ charge densities $\rho(r)$ are not available in Ref. \onlinecite{zbl}. We tested this approximation using cations of elements in the previous row of the periodic table where actual ZBL charge densities are available, viz. Nd, Pm and Sm. As can be seen from figure \ref{fig:f1}, the approximation satisfactorily captures the electronic shell structure of $4\pi r^2 \rho(r)$, which is the quantity of interest to us. Note that we have removed altogether any splines for cation-cation interactions. The downhill simplex method of Nelder-Mead was then used to carry out the potential fitting\cite{amoeba}. The fitting involved minimizing an objective function equal to the sum of the squares of the differences between the \textit{ab initio}/experimental data (weighted since they denote different quantities) and that predicted by the potential for all the classes of data as detailed above. GULP was used for energy calculations and for atomic-positions optimization\cite{gulp}.

Figure \ref{fig:f2} shows the quality of fit for the PuO$_{2}$ equation of states and single atom perturbation data, while figure \ref{fig:f3} shows the same for NpO$_{2}$. Table \ref{table:defect} shows the defect formation energies as obtained by us in the GGA+\textit{U} calculations, along with the corresponding values from the current potential and from the previous potentials published for these systems. We excluded the cation defect formation energies entirely from the fitting objective function. This can be justified by considering that (i) these energies as per \textit{ab initio} are already very high - upwards of 12 eV; (ii) from the case\cite{Tiwary} of UO$_{2}$, it is expected that \textit{ab initio} actually underestimates these energies, and thus they are even less likely to form; and (iii) these (Pu and Np) are the minority cations. It has been argued\cite{chartier} though that Uranium Frenkel pairs and Schottky trios might play an important role in the diffusion of noble gas impurities formed after fission - as such, our library of potentials does provide a much better match for the Uranium Frenkel pair and Schottky trio formation energy since it is based on the potentials in Ref. \onlinecite{Tiwary}.

The potentials so obtained are plotted in Figure \ref{fig:f4}, while the fitted coefficients are detailed in Table \ref{table:coefficients}. Note that since there was no spline in any cation-cation interaction (see Equation~\eqref{eq:eq1}), they do not find a mention in the above list. The aforementioned 5$^{th}$ order polynomial is uniquely determined by the provided cutoffs and potentials. The detailed potentials are available as a GULP library file.

\begin{table}[htp]
\caption{ Coefficients of fitted potentials}
   \centering
\begin{tabular}{c | c | c }
\hline\hline

\hline
       &
       PuO$_2$&
       NpO$_2$\\
       
       \hline
            A (eV) &
       597.304 &
      597.605 \\

       \hline
            $\rho$ ({\AA}) &
      0.475712 &
      0.484948 \\
      
           \hline
            B (eV{\AA}$^6$)&
       0.31187 &
      0.31187 \\
           
              \hline
         C (eV/{\AA}$^5$)&
       0.0003375 &
      -0.0735556 \\
      
             \hline
         D (eV/{\AA}$^4$) &
       0.029085 &
      0.048972 \\
      
             \hline
         r$_1$ ({\AA}) &
       1.42 &
      1.17 \\
      
             \hline
         r$_2$ ({\AA}) &
       1.7 &
      1.7 \\
      
             \hline
         r$_3$ ({\AA}) &
       2.85 &
      2.94 \\
      
\end{tabular}
\label {table:coefficients}
\end{table}

\begin{figure}[htp]
\centering
\includegraphics[width=90mm]{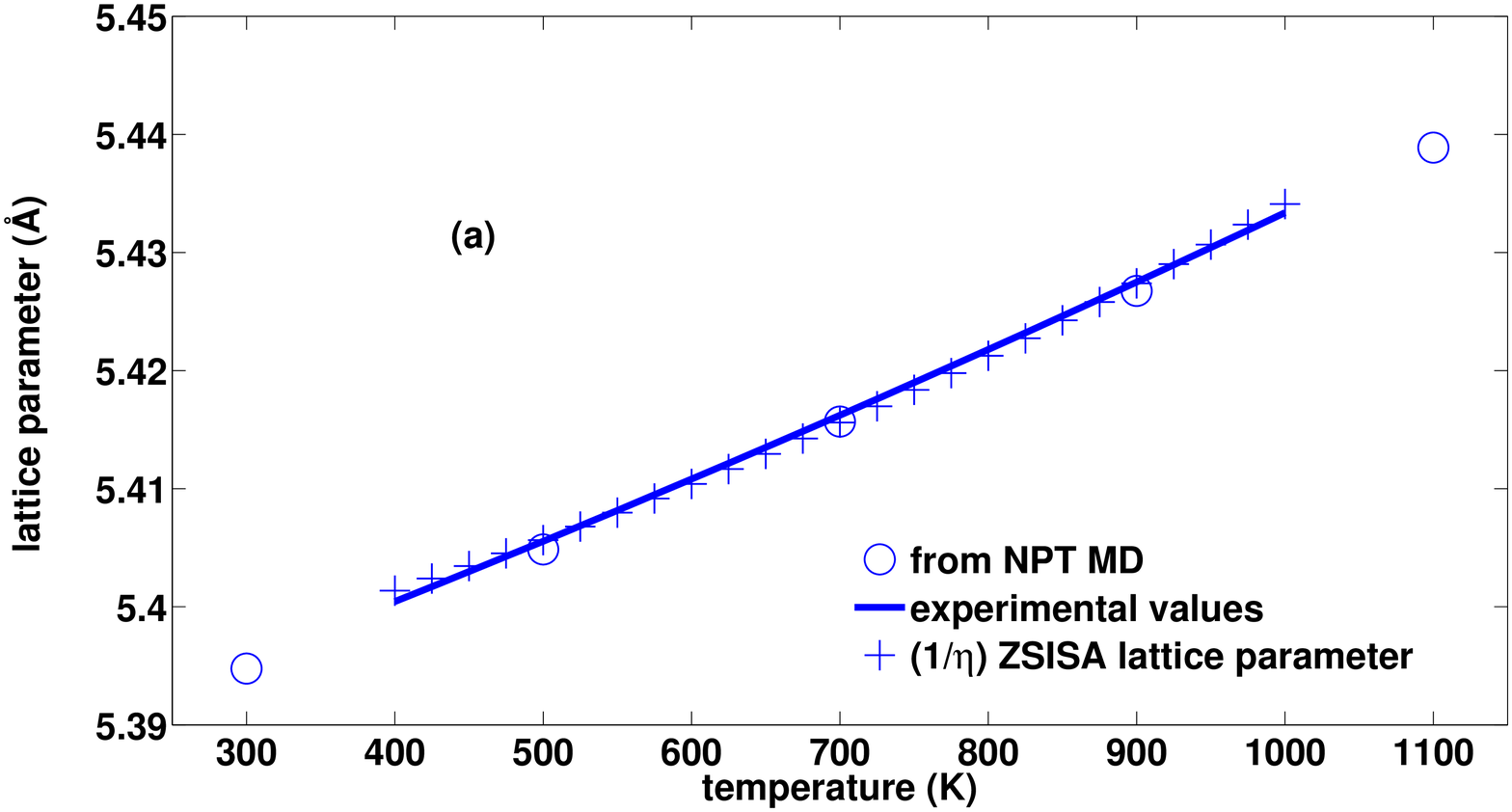}
\includegraphics[width=90mm]{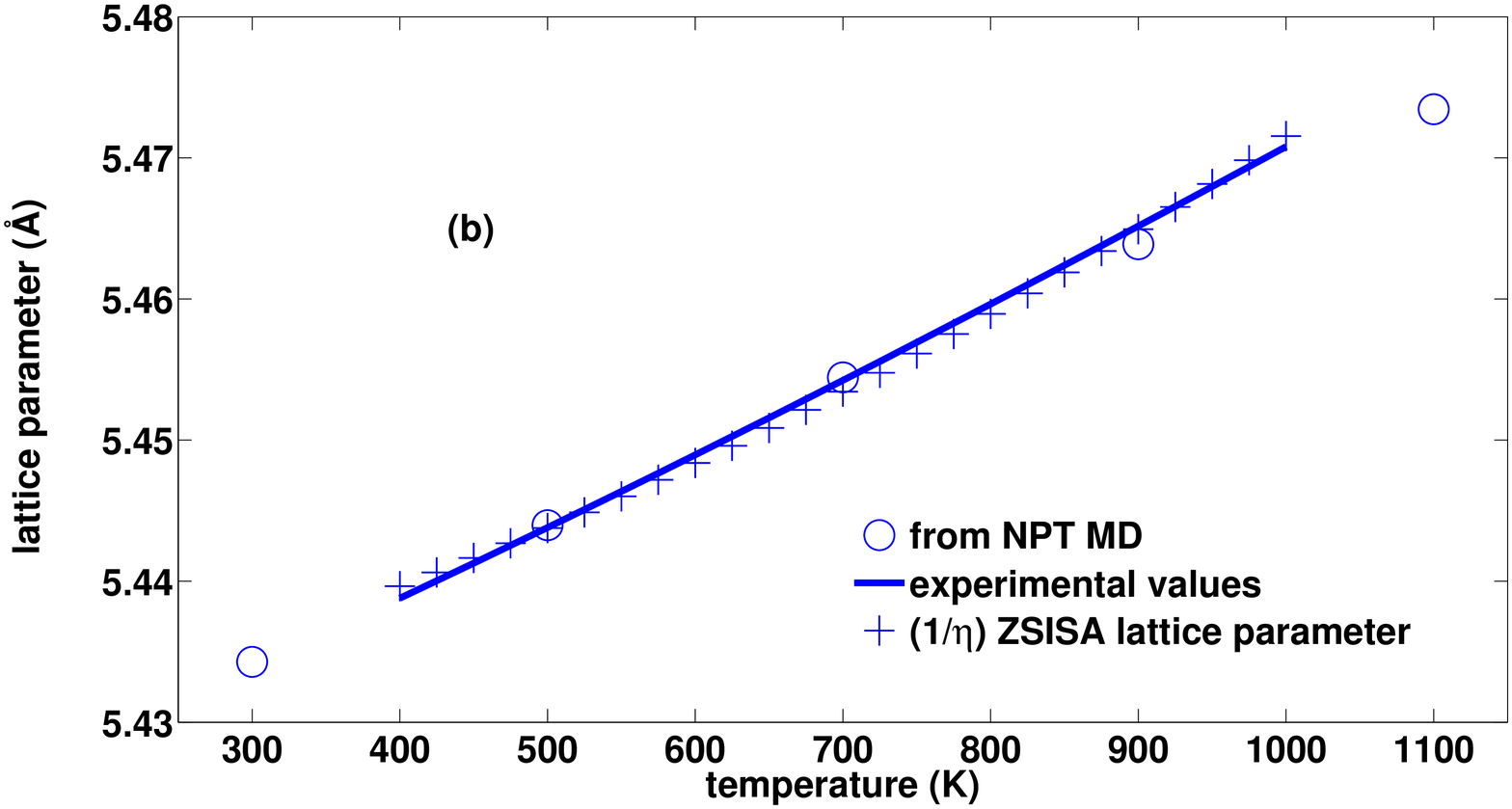}
  \caption[7]
{Lattice parameter at various temperatures for (a) PuO$_2$ and (b) NpO$_2$. Straight lines are the experimental values\cite{latparam} valid between 400 and 1000 K, while circles denote values obtained from MD simulations using current potentials. Plus signs represent (1/$\eta$) times the experimental values actually used in fitting to account for the observation that ZSISA slightly overestimates the MD lattice parameters. Details of calculation of this adjustment factor $\eta$ (equaling 1.0006 and 1.0008 for PuO$_2$ and NpO$_2$ respectively) can be found in the text.}
\label{fig:f7}
\end{figure}

\begin{figure}[htp]
\centering
\includegraphics[width=90mm]{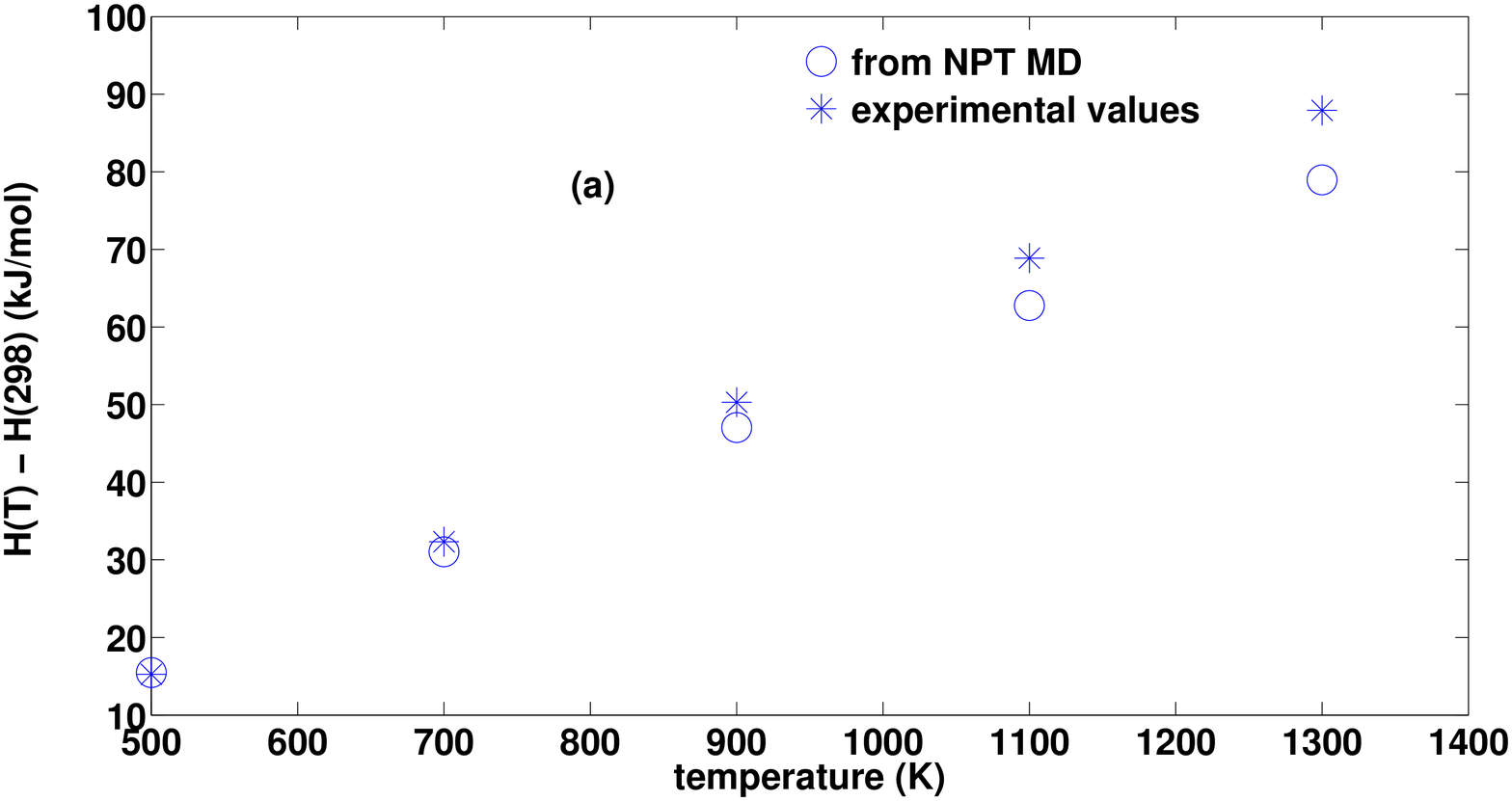}
\includegraphics[width=90mm]{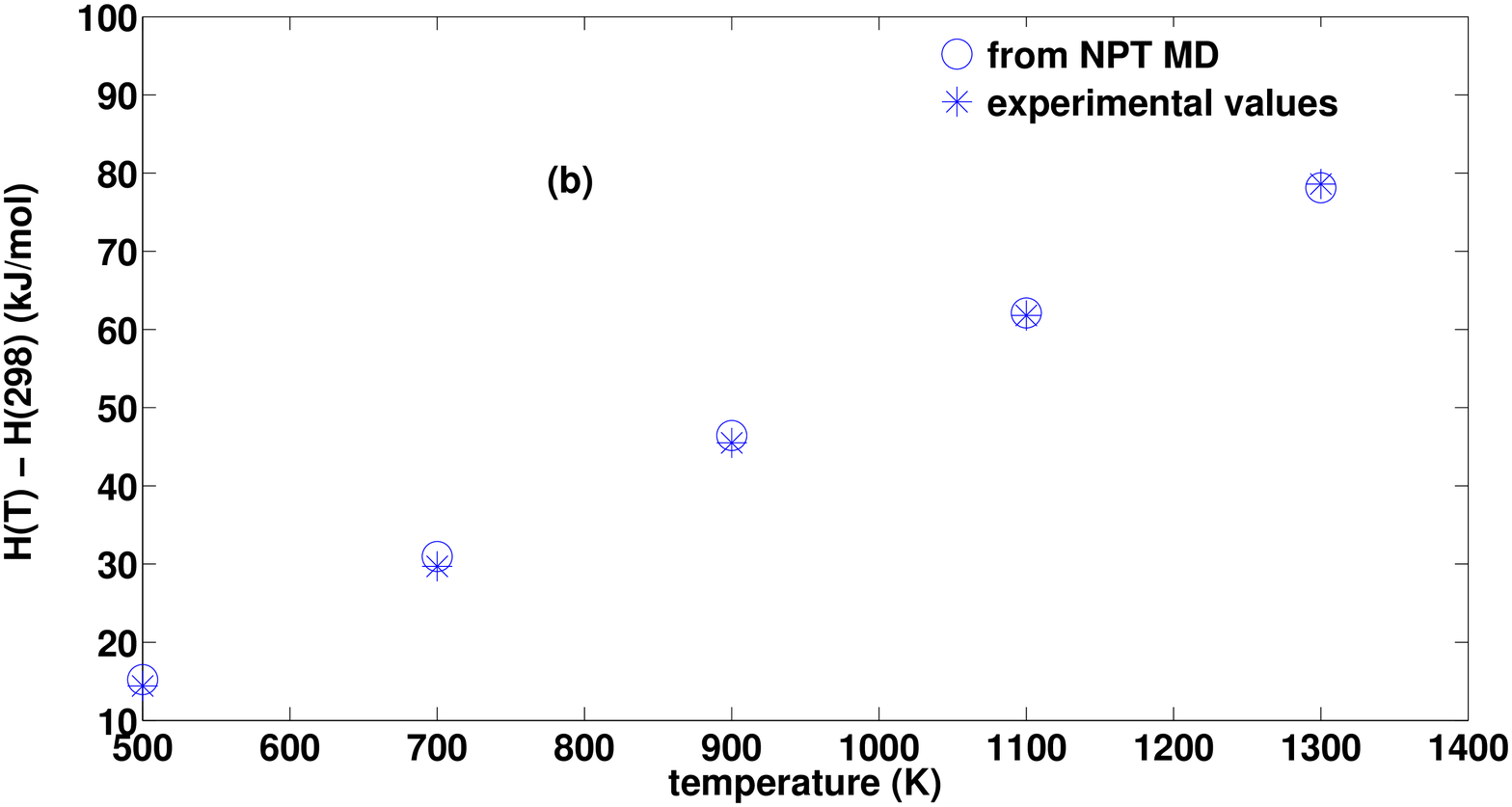}
  \caption[8]
{Enthalpy at various temperatures (relative to room temperature enthalpy) for (a) PuO$_2$ and (b) NpO$_2$. The circles denote values from NPT MD (predicted and not fitted values) while the asterisks are the known experimental values\cite{heatcap}.}
\label{fig:f8}
\end{figure}

The performance of the potential against the validation data, i.e., equation of states for oxides of U$_{31}$Pu, U$_{30}$Pu${_2}$, U$_{31}$Np and U$_{30}$Np${_2}$ can be seen from figures \ref{fig:f5} and \ref{fig:f6}. The match is satisfactory and interestingly it improves with more Pu or Np content in respective cases.

The generated potentials were  verified through NPT MD simulations on 3$\times$3$\times$3 unit cells (324 ions). The system was equilibrated for 10 ps while production runs were carried out for 100 ps with time steps between 0.001 and 0.0005 ps (depending on temperature). Apart from the lattice parameter, we also considered the enthalpy as a function of the temperature.

Figure \ref{fig:f7} compares the lattice parameter as obtained from the MD simulations with experimental values for PuO$_{2}$ and NpO$_{2}$\cite{latparam}. Figure \ref{fig:f7} also shows the corresponding ZSISA values as obtained from the potentials. The over-estimation adjustment factor $\eta$ used on the ZSISA values can be seen here. After this adjustment to ZSISA, the match for the lattice parameters between NPT MD and experiments is excellent. The quality of the enthalpy values compared between experiments\cite{heatcap} and those predicted from NPT MD with current potential is also very good (see figure \ref{fig:f8}).

To summarize, we have developed interatomic potentials for the Mixed Oxide fuel system (U,Pu,Np)O$_2$ by fitting to an extensive \textit{ab initio} database and to available experimental observations using a formalism that has been shown to be capable of dealing in a self-contained manner with conditions ranging from thermodynamic equilibrium to very high energy collisions relevant for fission events. The potentials capture known experimental measurements on these oxides as well as a rich database of \textit{ab initio} GGA+\textit{U} results. The applicability of these potentials in scenarios not included in the fitting is also explicitly demonstrated.

This research was supported by the US National Science Foundation through TeraGrid resources provided by NCSA under grant DMR050013N, through the U.S. Department of Energy, National Energy Research Initiative for Consortia (NERI-C) grant DE-FG07-07ID14893, and through the Materials Design Institute, Los Alamos National Laboratory contract 75782-001-09.



\end{document}